\documentclass{sigchi}

\CopyrightYear{2020}
\setcopyright{acmcopyright}
\doi{https://doi.org/10.1145/3313831.3376862}
\isbn{978-1-4503-6708-0/20/04}
\conferenceinfo{CHI'20,}{April  25--30, 2020, Honolulu, HI, USA}
\acmPrice{\$15.00}

\usepackage{balance}       
\usepackage{graphics}      
\usepackage[T1]{fontenc}   
\usepackage{txfonts}
\usepackage{mathptmx}
\usepackage[pdflang={en-US},pdftex]{hyperref}
\usepackage{color}
\usepackage{booktabs}
\usepackage{textcomp} 

\usepackage{microtype}        
\usepackage{ccicons}          

\usepackage{todonotes}

\def\plaintitle{Understanding the Use of Crisis Informatics Technology among Older Adults}

\def\emptyauthor{}
\def\plainkeywords{Crisis informatics; older adults; emergencies; human values.}

\makeatletter
\def\url@leostyle{%
  \@ifundefined{selectfont}{
    \def\UrlFont{\sf}
  }{
    \def\UrlFont{\small\bf\ttfamily}
  }}
\makeatother
\def\pprw{8.5in}
\def\pprh{11in}

\definecolor{linkColor}{RGB}{6,125,233}
\hypersetup{%
  pdftitle={\plaintitle},
  pdfauthor={\emptyauthor},
  pdfkeywords={\plainkeywords},
  pdfdisplaydoctitle=true, 
  bookmarksnumbered,
  pdfstartview={FitH},
  colorlinks,
  citecolor=black,
  filecolor=black,
  linkcolor=black,
  urlcolor=linkColor,
  breaklinks=true,
  hypertexnames=false
}

\begin{document}

\title{\plaintitle}


\numberofauthors{1}
\author{
    \alignauthor{Yixuan Zhang, Nurul Suhaimi, Rana Azghandi,  Mary Amulya Joseph,
    \\Miso Kim, Jacqueline Griffin, Andrea G. Parker\\}
    \affaddr{Northeastern University\\}
    \email{(zhang.yixua, suhaimi.n, azghandi.r, joseph.mary, m.kim, ja.griffin, a.parker)@northeastern.edu}
}

\maketitle
\begin{abstract}
Mass emergencies increasingly pose significant threats to human life, with a disproportionate burden being incurred by older adults. Research has explored how mobile technology can mitigate the effects of mass emergencies. However, less work has examined how mobile technologies support older adults during emergencies, considering their unique needs. To address this research gap, we interviewed 16 older adults who had recent experience with an emergency evacuation to understand the perceived value of using mobile technology during emergencies. We found that there was a lack of awareness and engagement with existing crisis apps. Our findings characterize the ways in which our participants did and did not feel crisis informatics tools address human values, including basic needs and esteem needs. We contribute an understanding of how older adults used mobile technology during emergencies and their perspectives on how well such tools address human values.  
\end{abstract}




\section{Introduction}

We are increasingly experiencing disasters and crises, including floods, hurricanes, and gas explosions that can have devastating effects on human life~\cite{eshghi2008disasters}. This is particularly true for vulnerable populations such as older adults who are more likely to suffer from mobility, sensory, and cognitive limitations that may impede decision making and taking action during crises~\cite{hoffman2008preparing}. 
Our work examines how Information and Communication Technologies (ICTs) can support older adults during natural-critical events that may threaten people's lives, across varying scales---for example, large and small, and short or no-notice events. In this paper, we use the following terms interchangeably: \textit{emergency} (an incident that threatens life, property or public health and safety ~\cite{blanchard2008guide}),  \textit{disaster} (a  disruption of a normal functioning system (e.g. power, water) that goes beyond the local capacity to respond~\cite{blanchard2008guide}, and \textit{crisis} (a situation in which people are unable to use of normal routine procedures and in which stress is created by sudden change~\cite{al2016understanding}). While these terms are defined distinctly, the concepts are clearly related---they are undesirable situations that present threats to the community and are often dealt with under uncertainty.
 
To mitigate the effects of disasters, prior work has examined the design of disaster assistance programs to specifically help older adults, including through establishing strong connections between older adults and available resources (e.g., governmental agencies, friends and family)~\cite{ngo2001disasters} and enhancing personal preparedness~\cite{federal2004you}.  
HCI researchers have also turned their attention to studying disasters and emergencies, with a focus on understanding the role of ICTs in times of crisis---a body of work known as \emph{crisis informatics}. Crisis informatics is as an interdisciplinary research area that investigates ``the interconnectedness of people, organizations, information and technology during crises''~\cite{hagar2010crisis}.

Prior work has argued for the great potential of using ICTs~\cite{perry2003internet}---and in particular social media---to facilitate emergency response activities such as information dissemination (e.g., emergency warnings and alerts)~\cite{jin2014examining,lindsay2011social}. However, while researchers have studied the role of crisis informatics tools in the general population, there has been less work examining how specific populations, such as older adults, use these technologies and the particular needs that may arise in these groups. And yet, technology adoption is growing amongst older adults. A 2017 study found that approximately 67\% of older adults in the United States use the Internet and 42\% own smartphones~\cite{pew2018mobilefact}. These trends, together with older adults' increased vulnerability during disasters, necessitate research that explicitly examines the needs and values of older populations during emergency situations. One critical perspective to understand is older adults' feelings about the services provided by emergency preparedness and response resources. For example, coercive measures taken during evacuations can threaten a person's sense of control and dignity~\cite{ngo2001disasters}, necessitating work that examines how to address these fundamental human rights, especially in populations where these values are particularly threatened, such as older adults. Despite the particular needs and strengths of ageing populations, little work has examined how emergency-focused technologies should be designed to address salient human values within this demographic (e.g., sense of control and dignity). 

To address this research gap,  our work aims to understand perceptions towards crisis apps among older adults.
Here we define \emph{crisis apps} as mobile apps that provide specific features and functions needed during crises, emergencies, or disasters~\cite{reuter2017katwarn}, and older adults as persons aged 65 and over~\cite{shih2018improving}.
Our work seeks to avoid the assumption that crisis apps for the general public will have the same efficacy amongst older adults. Such assumptions can be harmful for vulnerable populations by generating systems that disproportionately benefit more advantaged groups~\cite{veinot2018good}. This paper reports on our investigation of the role technology can play in addressing critical human needs, including basic needs (e.g., safety) and esteem needs (e.g., dignity), for older adults during emergency situations. We conducted a qualitative study with 16 older adults who had recent experience with an emergency evacuation. Our work is guided by the following overarching research questions: 
\textbf{RQ1:} How have older adults used crisis apps, which seek to provide help during emergencies, and what are their perspectives on using such tools in the future? 
\textbf{RQ2:} To what extent do existing crisis apps address older adults' values during emergency situations?

Our findings help characterize older adults' experiences with and perspectives on crisis apps that connect individuals with entities providing informational and instrumental resources during emergency situations. We found that there was a lack of awareness of and engagement with these crisis apps amongst the participants in our study, though most of them acknowledged the usefulness of crisis informatics technology. Our findings also highlight the role of community organizations in providing exposure to crisis informatics technology amongst older adults. Moreover, our findings characterize how well participants felt existing systems support human values (i.e., basic safety needs, and a sense of control and dignity).  

In this work, we contribute: 
1) an examination of the intersection of the vulnerability of older adults during emergencies and the role of crisis informatics, 
2) previously-unavailable knowledge regarding older adults’ perceptions towards crisis apps, including perspectives on how well such tools address their basic and esteem needs (which little prior work has examined), and
3) new recommendations for the design and evaluation of crisis apps for older adults (e.g., leveraging community-based organizations to increase awareness and engagement). 
 
\section{Related Work} 

\subsection{Crisis Informatics}

Crisis informatics is a field that combines theory, methods, and knowledge from the social and computing sciences to study how ICTs can support effective preparation for, response to, and recovery from disasters~\cite{hagar2010crisis, pipek2014crisis}. An important component of crisis informatics is examining how ICTs can support collaboration amongst individuals, emergency responders, and government agencies during emergencies~\cite{pipek2014crisis, sun2014enabling}. Prior research has documented the great potential of ICTs, particularly social media, in helping individuals, government agencies, and non-governmental organizations (e.g., Red Cross, etc.) to make better decisions in disaster scenarios and to mitigate the impact of disasters on communities~\cite{apuke2018social, gambura2018two, gray2018supporting}.  

Government authorities at the federal, state, and local levels have used dedicated apps and social media accounts to disseminate information during emergencies (e.g., alert messages, updated crisis status, and evacuation orders)~\cite{reuter2018social}. This top-down information flow plays a critical role in crisis informatics given that governmental agencies may convey a sense of authority and trust amongst individuals~\cite{steelman2015information}. Of course, trust in governmental entities does not exist in all societies or even across subpopulations and, as such, there is a need for research examining the factors that impact individuals' trust in crisis informatics systems. For example, in times of crisis, the inability to verify information sources can serve as a barrier to acceptance of this information~\cite{starbird2010chatter, steelman2015information, tapia2014good}. Beyond issues of trust, another challenge with top-down information channels is the unreliability that can occur due to potential system failures during disasters and the communication latency of official information~\cite{winerman2009social}. Additionally, the usefulness of some messages sent by governmental authorities has been questioned~\cite{steelman2015information}. Our work seeks to expand knowledge regarding the challenges that arise when ICTs are used to disseminate crisis information in a top-down manner, by investigating older adults' perspectives on such messaging channels (e.g., those enabled by disaster alerting systems).
 
While top-down information dissemination is a more traditional approach during disasters, the rise of social media platforms such as Facebook and Twitter have enabled rapid information propagation and collaboration amongst individuals in a bottom-up fashion~\cite{palen2009crisis, white2016social}. For example, Vieweg and colleagues examined the use of Facebook during and after the 2007 Virginia Tech shooting~\cite{vieweg2008collective} and examined how distributed, decentralized problem-solving approaches (e.g., Facebook groups) can help provide socially-produced accurate information in times of crisis. Gray~\cite{gray2018supporting} found that the use of social media helped disaster managers and reduced communities' risk.  Prior work has also shown that social media can be used to mitigate the impact of disasters, but that challenges can arise in the use of such tools, such as threats to information credibility and reliability ~\cite{reuter2018social}. Yet, there exists little insight into how older adults perceive bottom-up messaging platforms such as Twitter and Facebook in times of crisis.

\subsection{The Use of ICTs among Older Adults}

As compared with younger populations, older adults have used various ICTs with less frequency~\cite{czaja2006factors}. Motor, sensory, and cognitive declines are commonly cited as barriers to older adults' adoption of ICTs~\cite{knowles2018older, rogers2005technology, vroman2015over}. Other barriers to the use of ICTs among older adults include a lack of interest due to their hesitations about its value~\cite{vroman2015over}, privacy and security concerns, and fear~\cite{knowles2018older}.  
However, more recently, mobile technology adoption has been climbing among older adults. According to the Pew Research Center, around 67\% of older adults use the Internet. Approximately 42\% older adults own smartphones, compared with 18\% in 2013~\cite{Anderson2017tech}.
Also, older adults have shown enthusiasm for using and co-creating technologies that meet their needs and requirements~\cite{knowles2018older}.

A growing body of work in the field of crisis informatics is studying the role of mobile technologies during disasters, given the increasing popularity of mobile devices [57]. Prior work has shown that platforms such as smartphones have been particularly effective in disaster response because they provide users with unparalleled access to crisis information~\cite{haddow2013disaster}. In addition, the use of GPS capabilities within mobile apps to mark oneself safe (e.g., Facebook Safety Check) can be particularly useful~\cite{collins2016communication}. 
 
To our knowledge, there has been minimal work in the crisis informatics domain that focuses on specific populations who are more vulnerable, and have distinctive needs and perspectives in relation to mobile and networked technology, such as older adults~\cite{iwasaki2013usability, kuerbis2017older, mitzner2010older}. A notable exception is work by Howard et al.~\cite{howard2017they}, which studied the communication expectations and preferences of vulnerable groups in emergency settings. Their findings highlighted the importance of examining the impact of age, context, relative disadvantage, disability, and accessibility to technology on emergency communication needs and preferences. While the authors examined attitudes toward various communication channels (e.g., television and radio) within a variety of vulnerable populations, they did not examine the perspectives of older adults on mobile and social computing platforms. 
Given that aging adults are one of the most vulnerable groups in disasters, often have unique needs and desires around ICTs~\cite{hope2014understanding, kuerbis2017older, mitzner2010older}  and continue to grow in their adoption of ICTs, our work examines how this population believes ICTs can provide support during crisis situations, and their concerns regarding such tools.
 
\subsection{Older Adults' Challenges in Times of Crisis} 

Disasters and crises can severely impact the health and well-being of older adults~\cite{ngo2001disasters}, as they can be more vulnerable due to impaired physical mobility, diminished sensory awareness, preexisting chronic health conditions (e.g., arthritis, hypertension, and respiratory ailments), and social and economic limitations~\cite{fernandez2002frail, supporting2019nih}. These varied challenges can pose a threat to older adults' safety during emergencies. Yet, safety is a fundamental human need (as depicted in Maslow's hierarchy of needs~\cite{mcleod2007maslow}), and, as such, protecting individuals from harm during emergencies---especially older adults who may be more vulnerable---is a top priority in the development and execution of emergency response efforts~\cite{fernandez2002frail}. To keep aging populations safe during disasters, entities such as emergency management agencies have created disaster preparedness and assistance programs to specifically help older adults\cite{perry2001facing}. These resources have sought to provide various forms of support, including enhancing personal preparedness~\cite{federal2004you} and establishing strong support between older adults and available resources within agencies~\cite{ngo2001disasters}.

In addition to satisfying basic human needs, such as safety, it is important to examine how well older adults can maintain a sense of control during disasters. Indeed, maintaining a sense of control is important for older adults' well-being since a reduced sense of control can negatively impact psychological and physical health~\cite{ferguson2010optimism, lachman1998sense}. Yet, during emergencies, older adults' sense of control is particularly threatened~\cite{barusch2011disaster, ngo2001disasters}. For example, a qualitative study of flood disasters by Tuohy and colleagues found that older adults experience difficulties in maintaining a sense of control during critical events~\cite{tuohy2012older}. Even though emergency events are considered to be low-control situations in general (where most decisions are made by officials), these events may further undermine older adults' sense of control if their ability to exercise their rights and make decisions is diminished. For example, during Hurricane Katrina, older adults were often forced to relocate, resulting in overwhelming feelings of stress and anxiety that compounded the trauma caused by the disaster~\cite{hrostowski2012five}. 

Moreover, older adults encounter challenges in maintaining a sense of dignity during crises. Attributes of dignity include maintenance of self-respect and maintenance of self-esteem~\cite{jacelon2004concept}. The Office of the High Commissioner for Human Rights (OHCHR) developed principles for older adults to promote dignity~\cite{ohchr1991}. The OHCHR indicated that older adults should be able to exercise their rights and freedoms when residing in shelters and should be fully respected with regard to their dignity, needs, and privacy during emergencies~\cite{ohchr1991}. Given these challenges, there is a critical need for research that examines how well technologies that seek to provide support during emergency situations can also help older adults maintain their sense of dignity. 

As the number of older adults increases in the next decade, it is important to understand the vulnerabilities and needs of this population during emergency situations. While much prior work has focused on how to address basic human needs (e.g., safety)~\cite{burnett2007rapid, cloyd2010catastrophic, rothman2007vulnerable, tuohy2014older}, this work has rarely examined how technology can support these basic needs in older adults, nor how well ICTs help this population maintain a sense of control and dignity (which are categorized as esteem needs in Maslow's framework~\cite{mcleod2007maslow}) during crises. Our work seeks to address these important research gaps.

\section{Methods} 

We conducted a qualitative study to explore the perceived value of using crisis apps amongst older adults. Our study consisted of two sessions of semi-structured interviews with a diary activity in between sessions. 

\subsection{Participant Recruitment and Overview}

Upon approval from the Institutional Review Board at our institution, we recruited 16 participants from a community center in an urban city in the Northeastern United States from February to April 2019. This community center has been used as a place of shelter during emergencies. For example, a gas explosion event occurred in the Northeast region of Massachusetts in September 2018. Due to overpressure in the gas distribution system~\cite{NTSB2018}, a series of explosions and fires occurred and destroyed as many as 80 homes and buildings, forcing thousands to evacuate~\cite{Korte2018USAToday}. 

Our inclusion criteria were that participants must be aged 65 years or older and able to speak and write in English. To ensure that participants were able to complete the various study components, adults with a diagnosed psychiatric, developmental or cognitive illness were not eligible to participate. We used the Allen Cognitive Level screen (ACLS)~\cite{allen2007manual} to determine participant eligibility.  
 
11 participants were female and 5 were male, with a median age of 76.5 years old (IQR = 10.5). All participants were retired, but three worked in volunteer positions after retirement. 12 out of 16 owned smartphones (Android or iPhone). This rate of smartphone adoption amongst our participants is higher than recent nationwide surveys of mobile phone ownership in older adults (approximately 42\% among adults 65 and older in 2016)~\cite{Anderson2017tech}. 
Participants reported that they used a variety of apps, such as messaging, weather forecast, entertainment, and crisis apps.
All participants were affected by the gas explosion emergency described at the beginning of this section, though to different degrees. Some experienced lasting and continued impacts from the gas explosions, such as needing to renovate affected areas of their homes and replace appliances (e.g., water heater, stove, etc.). Others were less significantly impacted by the crisis, for example, evacuating for a few nights or staying in their homes without gas and electricity for a couple of days.   

\subsection{Study Design}

\subsubsection{Session 1}

In our first session, participants completed a demographic survey and a semi-structured interview, in which we collected information regarding their experiences during the gas explosion, specifically how they received the information about the crisis, their responses to the information, and their evacuation process. We also asked participants to describe any prior use of crisis apps. This session lasted approximately 40 minutes. 

\begin{figure}
 \includegraphics[width= 1.0\columnwidth]{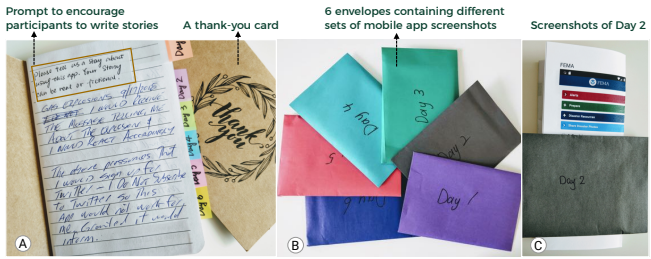}
 \caption{An example of kits being used in our exploratory design study}  \label{fig:culturalProbe}
\end{figure}

\begin{table}
  \centering
  \begin{tabular}{p{2.2cm} p{5.8cm}}
      {\textit{App}}
    & {\textit{Key Features}} \\
    \midrule
    Twitter Alerts~\cite{twitter} &  Get alerts and updates through FEMA Twitter account;  \newline
                                        Notify others with Tweets or direct messages \\
    FEMA~\cite{FEMAapp}  &  Get alerts/ updates and emergency tips;  \newline
                            Share disaster photos together with others \\
    iPhone SOS~\cite{iPhoneSOS} &  Send emergency contacts messages with current location \\
    Facebook~\cite{facebook} & Notify others about the crisis; \newline
                                        Stay updated with relevant information from different sources; \newline
                                        Mark oneself safe to reassure friends and family; \newline
                                        Give and find help with resources like supplies and shelters \\
    MEMA~\cite{MEMAapp} & Get up-to-date public safety information to keep informed in Massachusetts  \\
    Smart911~\cite{smart911} & Get alerts from the national and local public agencies; \newline
                                        A safety profile includes information about household  \\
    \bottomrule
  \end{tabular}
  \caption{Selected apps that seek to support people during emergencies}~\label{tab:selected_apps}
\end{table}

\textit{Exploratory Diary Activity:}
At the end of the first session participants were given diary kits to complete before the next session, to help them begin reflecting on how useful they feel ICTs can be in times of crisis. Each kit contained 6 days of activities; we chose this length of time to balance our goals of helping participants engage in meaningful reflection and being considerate of their time. We hoped that by creating this at-home activity kit, participants would be given the space to reflect upon their attitudes towards various existing systems designed for disaster scenarios, and that such reflection would help drive richer conversations during the subsequent interviews.
 
The diary kits (as shown in \autoref{fig:culturalProbe}) contained two primary components: 1) six envelopes with screenshots of mobile apps that seek to support people during emergencies and 2) a hand-sized journal. Each of the six envelopes contained screenshots of one of the selected apps. We used screenshots from the Google Play and Apple App Store and extracted descriptive text from these stores to explain the features and functionality of the selected apps. We also included a thank-you card in a seventh envelope to enhance user engagement. 
 
In the journal, we used labels with assorted colors to differentiate each day of the diary period, from Day 1 to Day 6. Each day, participants were asked to open one envelope, review the app screenshots in the envelope, and reflect upon the app in their journal. This reflection was guided by the prompt: ``Please tell us a story about using this app. Your story can be real or fictional''. Through this storytelling, participants would begin to reflect upon how and why they may, or may not, use the various crisis apps. 

In addition, participants were asked three questions about the app: How would using this app impact your 1) \emph{sense of control}, 2) \emph{sense of safety}, and 3) \emph{sense of dignity (self-respect and pride)} during an emergency? 
Participants could answer each question by choosing a rating from 1 (not at all) to 5 (a lot). We designed this exploratory diary activity to help springboard conversations in Session 2 around how older adults currently use crisis apps, and their perspectives on to what extent they would (or would not) use such tools in the future.

\textit{App Selection for the Diary Activity:}
We considered multiple factors when selecting apps for the diary kits. First, we considered crisis apps that were likely to be used by our participants; these apps included those adopted by the towns participants lived in and those recommended to us through informal discussions with administrative staff at the recruitment site. Second, we studied systematic reviews on mobile applications in crisis informatics literature including Tan et al.'s discussion of 106 apps~\cite{tan2017mobile}. We reviewed each of the 106 apps and selected a subset of mobile apps that met the following criteria: 1) available in English, 2) free to download and use, and 3) updated within a year of our review process. Lastly, we chose a set of apps that covered the major purposes of crisis apps, such as alerting and information provision, notification, crowdsourcing, and collaboration~\cite{tan2017mobile}.
With these considerations in mind, we selected the following apps, which include both top-down and bottom-up messaging channels: Twitter Alerts, FEMA, iPhone SOS, Facebook Crisis Response, MEMA, and Smart911 (see \autoref{tab:selected_apps}). 

\subsubsection{Session 2}

The second session was scheduled one or two weeks after the first session so that participants would still have the diary exercise fresh in their memories. This session was an in-depth semi-structured interview guided by our initial analysis of the interview data from the first session and the completed diary. Participants were asked to talk through their journals at the beginning of the session. We asked follow-up questions related to the written stories and the reasons for the ratings they gave for each app. This session lasted approximately 60 minutes. Participants received \$30 USD at the end of each of the two sessions.

\subsection{Data Analysis} 

Our research team wrote memos throughout the data collection and analysis period regarding important points that participants raised and topics that needed to be followed-up on in future interviews. All interview sessions were audio-recorded and transcribed verbatim. All stories written as part of the diary activities were digitized. We adopted the General Inductive Approach~\cite{thomas2006general}, where we conducted a close reading of the transcripts to get an initial understanding of the concepts arising in the data, carefully analyzed the transcripts (e.g., asking questions throughout the data), creating codes to label concepts in the data, and clustered related low-level codes to arrive at higher-level themes.
All authors discussed the emerging themes regularly to refine them  throughout the analysis process. 

\section{Findings}

Our findings provide insight into older adults' perspectives on the ways in which mobile technologies can be supportive in times of crisis, as well as their concerns with such systems. Through our analysis of the interview and diary data, we examined the role of crisis apps that seek to support communication, information seeking and dissemination, as well as the extent to which these tools support values that are particularly salient during emergency situations: safety, control, and dignity.  

Our participants described receiving emergency evacuation alerts from a variety of information sources. Half of our participants (i.e., P02, P04, P08, P09, P10, P12, P14, P15) first heard about the gas explosion and emergency evacuation order through interpersonal channels, such as phone calls from family and friends. Others (i.e., P03, P05, P06, P11, P13) first heard about it through the Reverse 911 system, which sends automated telephone calls that are managed by local emergency management agencies~\cite{strawderman2012reverse}.  Individuals were told to turn off the gas and get out of the house immediately. There were also participants who first learned about the crisis through mobile crisis app channels (i.e., P01, P06, P07).

In the following sections we describe our findings, which highlight the lack of awareness of and engagement with crisis apps amongst our older adult participants. We then discuss the ways in which participants felt existing platforms support and neglect the values of safety, control, and dignity in times of emergency.

\subsection{Awareness of, and Engagement with, Existing Tools}

Alerting individuals about emergency situations, directing them to take appropriate action, and providing information are some of the most common functions of technologies designed to provide support during emergencies. Such tools are primarily used for disseminating authorized information before and during disasters~\cite{tan2017mobile}, and include dedicated tools with top-down channels (e.g., FEMA, MEMA, CodeRED, and Smart911), as well as social media platforms as bottom-up channels that integrate disaster and crisis functionalities (e.g., Twitter Alerts, Facebook Crisis Response). In this section, we first report findings regarding the awareness of and engagement with current tools, and then we focus on the perceived usefulness of these tools.

\subsubsection{Lack of Awareness and Engagement}

We found that most participants were not aware of mobile crisis apps. Only four participants said they had heard of such technologies (e.g., FEMA, MEMA, and CodeRED), and three mentioned that they used such tools before the gas explosion event. These three participants (P01, P06, P07) said that during the gas explosion event they first received alerts to evacuate their homes from CodeRED, a mobile app that delivers public alert messages to subscribers within their local geographic area. CodeRED has been adopted and officially used by one of the towns affected by the gas explosion, and thus local governmental officials can notify individuals of an emergency or important situation. Even though CodeRED has been used officially by their town for crisis information communication and management, most participants (n=13) had not heard of it at the time of our interviews. Both P06 and P07 first learned about CodeRED from a public information session in a community center. P01 first learned about it while he was doing local volunteer work:
\begin{quote}
    P01: I used to work with the police department, volunteer work. So, I knew about it from the people I interact with here, \textit{``Hey we're (the police officers) coming up with this CodeRED.''} Cause a part of [organization] is a volunteer organization that connects the public safety offices. 
\end{quote}  
Thus, for those who used CodeRED, their participation in local organizations was central to their exposure to the system through which they first learned about the gas explosion. Leveraging community-based settings to introduce crisis apps may be particularly important given that these platforms are typically used infrequently, and thus individuals may have little opportunity to become acquainted with them. P02 wrote in her diary, for example, that she would need advance exposure to crisis apps:
\begin{quote}
    P02: Everything that was there I had to learn well before an emergency came along. I wouldn't be able to use any of this stuff during an emergency. 
\end{quote}
P02's comments, as well as the overall lack of crisis app awareness amongst our participants, highlight a key factor that may impede the effectiveness of these platforms amongst older adults: a lack of prior exposure to crisis apps. Indeed, P02 further commented on prior experience with technology:
\begin{quote}
    P02: Most people my age, late 60s, we are all saying to ourselves, \textit{``Our head is too full. We have no more room to learn something new''}... You guys have grown up with technology. You don't know any difference, but I didn't grow up with technology...
\end{quote}
P02's quote indicated she was resistant to adopt new technologies. Similar to P02, P16 also believed that technology was too much for her:
\begin{quote}
    P16: I think a lot of older people won't because the technology is just beyond us...
\end{quote}
This can be especially true given that emergencies are high stress situations, which can further impede one's ability to learn something new.  

Also, community organizations appear to be a promising venue for providing such exposure as participation in such institutions was common amongst our participants. For example, several were engaged in community volunteer work following retirement (e.g., P01, P06, P07, P08), including volunteering at a community center that was used as an emergency shelter during the gas explosion event.  

In addition to the lack of awareness of crisis management tools, we also found that despite growing interest in using social media for crisis information communication~\cite{simon2015socializing}, most of our participants did not engage with social media. Since most participants ($n=14$) do not use social media platforms in their daily life, they were not able to use Twitter Alerts that require users to sign into Twitter to subscribe to alerts from emergency authorities. For example, P01 explained that Twitter Alerts would not work for her:  
\begin{quote}
    P01: I do not subscribe to Twitter, so this app would not work for me, granted it would inform us.
\end{quote}

Some participants (e.g., P05, P01) also believed that other people their age do not use social media either. For example, P05 told us that \textit{``I do not have a Twitter, nor do I know anyone else in my age group that has a Twitter''}. Similarly, P09 also indicated that \textit{``I don't know one person of my contemporaries that does Twitter''}. 

The two participants who used social media indicated they only checked Facebook sometimes. Prior work has documented the significant difference in social media use by age~\cite{Aaron2018social}. According to a 2018 survey of social media use amongst Americans, approximately 88\% of 18- to 29-year-olds use some form of social media~\cite{Aaron2018social}. This share drops to 37\% among adults 65 years old and older, and only 14\% of older adults use Twitter~\cite{Aaron2018social}. Clearly, the lack of engagement and familiarity with social media makes it challenging for older adults to take full advantage of these platforms during emergencies.
  

Despite the lack of awareness of and low engagement with crisis apps, most participants acknowledged the usefulness of technology-driven alerts during emergencies. One participant (P04) wrote a story during the diary activity about Twitter Alerts that was based on her real experience:
\begin{quote}
    P04: During the winter of 1978, I was doing a home visit with a client... When I left her home, the wind and storm was so bad I had to hold on to a light post to prevent falling. I returned to the office and learned others were going home. If I'd had an alert system and if cell phones had existed back then, I could have learned it was a major storm and a state of emergency had been declared. Updated alerts would help understand current status of situation as situation changed. This would help me better respond as conditions change and know how to prepare... However, I use texts and emails but I don't use Twitter. I might do so if service for emergency notifications were only available that way or if I understood the advantage of that system.
\end{quote} 
P04's story, on the one hand, indicates that alerts could have helped her stay informed and prepare for the emergency. On the other hand, in addition to the lack of engagement with current social media tools (i.e., Twitter), P04's quote gives valuable insight that crisis apps may be only useful when people understand and are aware of their functionality. Some other participants (e.g., P01, P03, P06, P07, P08, P16) also believed that such alert messages could help them stay on top of crisis situations and react properly.

\subsection{Privacy and Information Security Concerns}
\label{sec:privacy}  

In addition to discussing their level of awareness of and engagement with current tools, our participants raised concerns about privacy and information security both in the interview sessions and the stories written in their diaries. For example, P04 explained her confidentiality concerns about Facebook's crisis support features. Facebook's Crisis Response contains a variety of features. For example, Safety Check aims to give users a sense of reassurance by allowing them to notify family members that they are safe during a crisis. Additional features include a platform for fundraising and providing or finding help during emergencies. P04 wrote a story about using these Facebook features, which was based on a real past experience that she had:
\begin{quote}
    P04: When a woman on the elevator with me was not able to reach her husband on his cell phone, she was very upset. If her husband had this app on his cell phone for Facebook Safety Check, he could have notified his wife he was safe. Connecting people to give and receive assistance is a benefit as well as fundraising and resource referral functions. I have some concerns about Facebook and confidentiality... I would prefer not to use Facebook for a lot of communication. I'm unsure how this would be better than texting loved ones. It publicized your situation and some people may be reluctant to have the whole community know about them... It seems like you'd need to sign on to Facebook, right? Yeah. I would prefer another system.
\end{quote}
P04's story shows how she acknowledged the value of using Facebook Crisis Response. However, the risk of publicizing an individual's personal information made participants like P04 lose trust in Facebook. We saw this hesitance to using social media tools like Facebook among many other participants (e.g., P01, P02, P05, P13, P16) who echoed P04's concerns. For example, P16 indicated that \textit{``because of the confidentiality question. I don't trust [Facebook]''}.

In fact, prior work has identified the privacy and information security risks of using social media during disasters. For example, information from social media can be misused for mischievous pranks or even acts of terrorism in times of crisis~\cite{conrado2016managing, lindsay2011social}. As prior work has reported~\cite{conrado2016managing, lindsay2011social}, we found that participants felt especially vulnerable to receiving misinformation due to their older adulthood status. Our findings expand upon this prior work by further highlighting how this sense of vulnerability in turn increased these concerns about using social media during emergency situations. Specifically, some participants were wary of crisis apps that might open older adults up to scams. For example, one participant (P03) talked about her concerns around using Facebook Crisis Response:
\begin{quote}
  P03: I think if it's generally something like this---that I'm safe--0that's one thing. If it gets into specifics, I think, by this point, I had started thinking... having some concerns about the amount of information being given out. That in a crisis situation that's also a time when there are people who can prey on people who are vulnerable. I was wondering how secure information is. I think especially the elderly people can be targets for scams. So, I guess had some concerns about that.
\end{quote}
P03's quote indicates that the degree of information being given out is an important consideration in times of crisis. She felt that simply marking herself as \textit{``I'm safe''} was acceptable, whereas revealing more specifics and details might bring up more concerns and uncertainty. 

Our findings suggest that the privacy and information security concerns around social media were not only coming from trust issues with the platforms themselves but are also associated with older adults' vulnerability to misinformation (e.g., spam) in times of crisis.  With regards to social media institutions, while they allow for connecting with trusted individuals and organizations, participants expressed concern and a lack of trust in how the information may be used by other organizations or individuals connected to these technology platforms.

\subsection{Seeking to Address Human Values} 

\begin{figure}
  \includegraphics[width= 1.0\columnwidth]{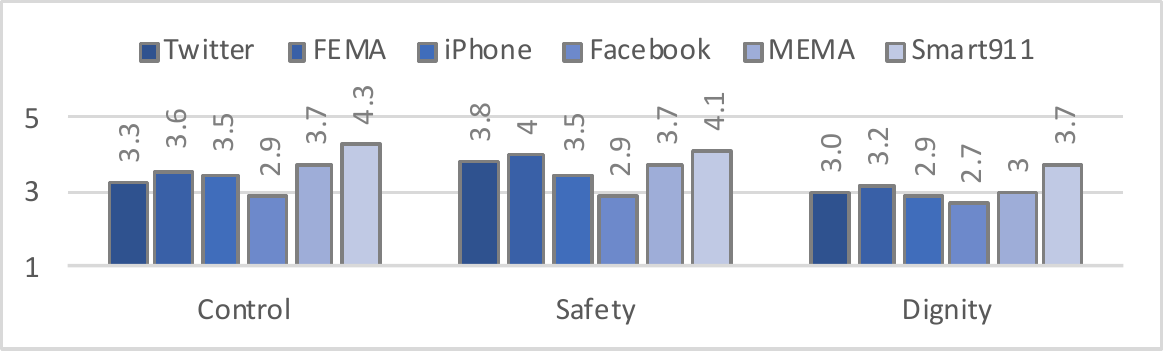} 
  \caption{Ratings of ``How would using this app impact your
\textit{sense of control/ safety/ dignity} during an emergency?''. 
Scores were provided by participants in the diaries, with a rating from 1 (not at all) to 5 (a lot).}
  \label{fig:rating_barchart}
\end{figure}

Our diary activities also revealed themes regarding how well existing tools support values that prior work has deemed salient in the context of emergency evacuation~\cite{ohchr1991, tuohy2012older}: \emph{sense of control, safety, and dignity}. Recall that for each day of the diary activity, participants were asked to review screenshots of a mobile app designed for emergency situations, and rate how they felt using this app would impact their sense of control, safety, and dignity during an emergency. This exploratory activity helped participants begin to reflect on how well they envisioned crisis apps aligning with their values. 
During our data analysis, we drew inspiration from Maslow's hierarchy of needs framework~\cite{mcleod2007maslow} as we unpacked the findings arising from the interview and diary data. Specifically, our findings describe participants' perspectives on how well crisis apps can meet more basic and fundamental safety needs (i.e., the need for protection and security) as compared with higher-order esteem needs (i.e., those associated with a feeling of respect and independence, such as dignity and a sense of control).  We use these concepts to help structure our resulting findings around participants' values.

We begin this section by reporting descriptive statistics summarizing our participants' ratings. As shown in Figure 2, Smart911 received the highest scores across all three value categories. Facebook was rated lowest in all categories---with mean scores under 3. These findings suggest that overall, participants did not feel that Facebook's Crisis Response features could provide them with a sense of control, safety, or dignity. Comparing how individual apps were rated across the three value categories, we found that each app scored highest in terms of its ability to address the more basic need of safety, but each scored lower in terms of the ability to address the higher-order values of control and dignity. 

All apps scored the worst in terms of their ability to provide a sense of dignity. All but one app (Smart911) received an average score of 3.2 or lower for dignity, reflecting that participants were ambivalent about the extent to which most apps could support this value. In the remainder of this section, we will discuss themes that arose across participants' reflections on how well the apps in the diary activity address basic needs (i.e., safety) and esteem needs (i.e., dignity and control).

\subsubsection{Supporting Connections to First Responders}

Our participants believed that being able to connect to first responders (e.g., emergency medical personnel, police officers, and firefighters) and provide them with more information could enable personalized treatment. This personalization would in turn provide reassurance, further enhancing their sense of control, safety, and dignity. In Smart911, a Safety Profile allows users to enter their personal or household information, including name, contact details (e.g., email, phone number), and medical information (e.g., health conditions, medications). When people call 9-1-1, their Safety Profile information is shared, allowing the 9-1-1 operators to recognize the caller's information~\cite{smart911}.  P07 discussed how the Safety Profile feature could benefit her during emergency situations:   
\begin{quote}
    P07: Many times in an emergency, our minds are so stressed that we can forget important medical conditions or other information that would be helpful to first responders. I would think that speaking to a person who is looking at our profile, therefore able to speak more personal to us, would be helpful to calm us down.
\end{quote} 
P07 believed that it would be beneficial for her to use apps like Smart911. During emergencies, it is not unusual for people to feel stressed and anxious. Enabling individuals to connect and share medical information with first responders can help them react more effectively to individuals' specific medical needs. As P07 said, this personalized care may be valuable in helping calm adults' nerves. Such a calming process can be instrumental in helping people regain a sense of control~\cite{alexander2001ambulance}.

Moreover, being able to connect and share information with first responders can help people feel safe. For example, P04 drew upon her past experience as a hospital social worker to write about the Smart911 app in her journal. She felt that being able to retrieve medical information in Smart911 could be very useful, especially for people with serious health challenges. P04 wrote in her journal: 
\begin{quote}
    P04: Particularly for people with complex medical conditions, Smart911 would facilitate medical providers having information needed. Patients would feel more safe knowing that even if they became unresponsive, medical care could be informed and appropriate specialists consulted... for assistance will be made available is reassuring and gives a sense of control... 
\end{quote}
P04's quote describes a scenario of patients obtaining a sense of safety by informing first responders during an emergency. P04 further added that sharing information with medical personnel would enable her to feel less vulnerable, because \textit{``emergency personnel could be contacted quickly and efficiently''}.

In summary, we found that our participants appreciated the functionality of informing first responders of their health conditions in advance so that  medical personnel can react accordingly and provide more personalized care. They believed such features could help them feel safe and maintain a sense of control.

\subsubsection{Ability to Initiate Action}

In addition to being able to connect with first responders, some participants (i.e., P01, P06) indicated that the ability to initiate action could increase their sense of safety, control, and dignity. For example, P06 appreciated the iPhone SOS emergency feature, which allows people to quickly call for help and alert emergency contacts. Pressing and holding the side button and either volume button will trigger the SOS emergency mode. Once entering SOS mode, the iPhone will send the user's real-time location to their emergency contacts. P06 explained how he felt about triggering the SOS emergency mode:
\begin{quote}
    P06: The interaction process leads towards a sense of response... many of these apps give you information. You say, \textit{``Oh, now what should I do now?''} And you open the app and it says, \textit{``Oh, you can go to a shelter here, or you can do something else.''} But that app [iPhone SOS] lets you do something that was interactive, so that you pushed a button and it sent your location to the 911 kind of people.
\end{quote}
In this case, P06 advocated the interactive feature of iPhone SOS that allows mutual communication between people who seek help and people who provide help in times of crisis. Being able to \textit{``push a button''} to send notice to first responders---a way to initiate action during emergency---may provide reassurance to individuals at that moment. Similarly, P01 liked the iPhone SOS function and highlighted how taking action during emergency situations impacted his sense of dignity (self-respect):   
\begin{quote}
    P01: [iPhone SOS] gives me the ability to call for help... So therefore, there's some level of self-respect, because I took the initiative to use what I had as a tool.
\end{quote} 

\subsubsection{Addressing Esteem Needs}

While iPhone SOS supported action in a way that P01 viewed as enabling a sense of dignity, participants were generally skeptical that the tools in the diary activity could facilitate a sense of dignity. Compared with the other values we assessed (i.e., control and safety), all apps were rated lowest in terms of their ability to provide a sense of dignity. This finding is partially explained by the fact that for some participants, dignity was not conceived of as a product of technological interaction. For example, P07 described dignity as more of a ``personal'' quality as opposed to something one gets from a technology. Similarly, P02 said \textit{``I don't look to technology for my dignity, that comes from inside me''}. 

At the same time, participants reflected in their journals about how technology can be useful for helping one maintain a sense dignity during emergency situations. For example, P03 recalled her past experience forgetting to bring medications with her. She further indicated that she appreciated the ability to connect to emergency personnel using the Smart911 app,

\begin{quote}
    P03: Being reassured that the information needed to keep you safe is available to emergency personnel is helpful. It could be embarrassing to have difficulty trying to think of all of your meds and dosages at a stressful time---would not help you feel dignified and secure.
\end{quote}
P03's concerns regarding ``embarrassment'' arising from forgetting medications ``at a stressful time'' echo prior work describing how medical conditions can lead to concerns around awkwardness and humiliation, particularly when it comes to asking for help during emergency situations~\cite{martin2000self, miller1992nature}. 

In addition to the moments of indignity in times of emergency, our participants (e.g., P16) also described how tools like Smart911 could help them maintain a sense of dignity and pride, particularly the Smart911 profile that allows users to keep medical notes. For example, P04 reflected on her past experience working with patients and said the following while reviewing her journal: 
\begin{quote}
    P04: Then dignity and pride, I said, ``People would be more free to travel and feel assured of receiving appropriate care. This would help them feel less anxious and pressured to provide complex and accurate medical information. Patients [that were impacted by the crisis] and family members need to be able to advocate for themselves and having this information to refer to would confirm the accuracy of their verbal reports and could enhance communication with medical providers.'' 
\end{quote}
P04 felt that being able to express and advocate for herself, ensure information accuracy, and communicate with first responders may help her maintain a sense of dignity. Indeed, being rushed and not being informed of medical treatment compromise one's sense of dignity~\cite{walsh2002nurses}.

We also found that many participants (e.g., P01, P03, P04, P06, P07, P08, P13, P16) saw dignity as being very much tied to other values (e.g., sense of control). For example, P16 described this interconnection of such values this way: 
\begin{quote}
    P16: ... especially when people are feeling so vulnerable with illness and uncertainty [during crisis]. Whatever you can do to enhance their sense of control and privacy for discussions or for physical exams or things. Those are all important in terms of treating people with dignity.
\end{quote}

As the findings in this section begin to demonstrate, the sense of control, safety, and dignity were closely linked as our participants described their perspectives on crisis apps. 
P01 discussed these relationships when he reflected upon his diary based on his prior experiences interacting with a variety of crisis apps: 
\begin{quote}
    P01: The more prepared you are, the safer you'd be. You have the ability to make your own decisions, so therefore, you're able to control your fate more, so a higher level of dignity associated. They are intertwined.  
\end{quote}
Our findings suggest that older adults in our study felt existing tools were best suited to help them meet the basic human need of safety, but poorly-suited to address esteem needs, particularly a sense of dignity. They further described the needs of safety, control, and dignity as very much interdependent. Thus, while emergency response entities are increasingly attempting to use ICTs to reach and help people during crisis scenarios, our work suggests that these entities may be more successful at empowering older adults if they support needs such as safety, control, and dignity in concert.
 
\section{Discussion}

Our findings characterize our participants' varied perspectives on crisis apps. Participants saw promise in various features, such as being able to connect with first responders, and tools that allow them to take action during a crisis. At the same time, participants were largely unfamiliar with tools that can provide help during emergencies and had concerns about the suitability of such platforms, for example, in terms of how accessible they were to an older demographic (e.g., as in the case of Twitter Alerts), and in terms of their ability to honor values such as control and dignity. In the remainder of this section, we build upon our findings to argue for future work that examines approaches to promoting awareness of and engagement with crisis informatics tools, as well as addressing human values as design goals. 

\subsection{Promoting Awareness and Engagement}

With the growing prevalence of social media use, these platforms have demonstrated great potential in supporting crisis communication and management~\cite{haddow2013disaster, xiao2015understanding}. During emergency situations, people prefer to use familiar tools that they have already used frequently before the onset of a crisis~\cite{haddow2013disaster}, such as Twitter and Facebook. However, older adults in our study did not use social media tools, and some went further to speculate that other older adults their age rarely do either. As mentioned previously, this limited social media engagement echoes established trends that older adults are less likely to utilize social networking platforms than younger populations~\cite{kuerbis2017older}. In addition to the low engagement with social media apps, most adults in our study reported that they were not aware of the currently-available crisis apps, including the CodeRED app that has been officially adopted by their town. Given the importance of being familiar with a tool prior to an emergency event~\cite{tan2017mobile}, the lack of awareness of and engagement with existing crisis informatics systems may create barriers to using such tools during emergencies. 

Our work highlights the importance of future research that investigates older adults' concerns and values around crisis informatics technologies and information sharing, how comfortable and engaged older adults are with such systems prior to and during emergency situations, how those norms and values evolve over time and may impact the use of crisis informatics systems, and how older adults' needs may align with or diverge from those of younger adults. Indeed, attending to these issues is crucial to avoid creating intervention-generated inequalities, that is, inequity in well-being outcomes that are generated because newly-created technologies are more accessible, usable, or effective in one population than another~\cite{hope2014understanding, kuerbis2017older}. In the case of systems like Twitter Alerts and Facebook Crisis Response for example, if these platforms are seen as more out of reach for older adults but easily adopted by younger populations, this can create a scenario in which younger adults have greater access to the informational, social and financial resources that these platforms offer during emergency situations. Such increased access may mean that younger adults are more able to protect themselves from the negative impacts of an emergency than older adults (who are already vulnerable in such situations, due to health challenges or limited financial resources of a fixed income). As such, there is a vital need to consider how crisis informatics tools can be made more accessible, useful, engaging, and effective for older adults.

Moreover, there is reason to believe that such systems can in fact be more effectively designed for older adults. Despite describing various shortcomings of existing tools that seek to support people during emergencies, most participants in our study still felt that these kinds of technologies could help them be more informed about and prepared for emergencies, as such, they were willing to consider using these technologies. These findings suggest a great opportunity for future work to create tools that better support older adults.
  
Among our participants who had prior experience with such apps, they all first learned about them through community-based volunteer experience or public information sessions held in local community centers. While this neighborhood engagement is to be expected given that we recruited participants through a community center, it is also reflective of prior research that has found that community volunteerism has been steadily increasing amongst older adults~\cite{haddow2013disaster}. In fact, in 2005 older adults were the most likely age group to volunteer 100 or more hours a year~\cite{haddow2013disaster}. 

Our findings highlight the value of leveraging community-based organizations to increase awareness of and engagement with new tools (e.g., crisis apps). First, given the vital role that neighborhood institutions such as community centers and fire departments play during emergency scenarios~\cite{chandra2013getting}, there is a great opportunity to explore how software tools can be designed to better connect older adults and neighborhood organizations. Such systems could enable older adults to communicate their needs to local organizations providing disaster response services, given that such proactive action was valued by participants in our study.  Second, designing  platforms that thoughtfully bootstrap technology learning communities might help connect older adults with varying degrees of prior exposure to crisis apps. Being introduced to and trained by people who they trust (e.g., family and friends, or staff at community centers that older adults are familiar with), might help reduce the anxiety of technology adoption and increase the self-efficacy of using the new technology~\cite{vroman2015over}. Correspondingly, our study suggests the value of designing sociotechnical systems that enable those who have interacted with crisis apps to share their experiences with and educate others (e.g., through legitimate peripheral participation~\cite{lave2001legitimate, lave1991situated}). Such innovations and broader work attempting to meet older adults' needs will require not only digital tools but also community partnerships to create services and develop relationships between institutions and community members that can support both esteem and basic needs. We suggest future work investigating the creation of such sociotechnical systems, particularly the challenges, assets, and affordances within different communities that may impact innovation.

\subsection{Addressing Human Values as Design Goals}

In this paper, we explored how well existing technologies address three key human values in times of crisis (i.e., control, safety, and dignity). Our findings suggest that while there are significant shortcomings, some existing systems may appropriately address the basic human need of safety in older adults. In particular, our participants believed the capability of connecting to first responders, being treated more personally, and the ability to initiate action could help them gain a sense of reassurance, safety, and control. 
For example, some participants valued the ability to initiate connection with first responders and emergency contacts (e.g., through iPhone SOS), as a way to maintain a sense of control and dignity.
However, care must be taken to investigate how sense of control and dignity can be impacted when technology fails to work as expected (e.g., when responders are delayed), or when trust in such tools are low. Work is especially needed to examine the perspectives of older adult subpopulations, and groups who may have lower trust in emergency response agencies due to issues like racism and discrimination. We encourage future work that examines the limits of technology in facilitating esteem needs---and opportunities for combining social with technological innovations to fully meet older adults' needs.
 
Our work also specifically highlighted how dignity was perceived as the least supported value by the existing platforms, and how basic needs (safety) and esteem needs are intertwined. In fact, there are many threats to older adults' sense of self-respect and pride in emergency scenarios, such as the embarrassment that one of our participants described at forgetting her medication list when communicating with a first responder. Identifying such threats to dignity and designing tools that mitigate these threats may be an important way of helping improve older adults' emotional well-being in times of crisis. 

Increased work is also needed to further examine to what extent current crisis apps address other human values (e.g., calmness, identity, and courtesy)~\cite{friedman2008value} in vulnerable populations such as older adults. While we have focused on the values of older adults with regard to existing systems, further research is needed to examine whether there is a need for fundamentally-different and new technology or if there are ways to adapt current technologies.  Can the same technology be used to address multiple populations, such as those with disabilities, and low socioeconomic status? Future work should also examine how challenges around addressing esteem needs (e.g., dignity) might be manifested differently or similarly in these varied vulnerable groups.

\section{Limitations}

As noted previously, our selection apps for the diary activity covers a small number of tools in the crisis informatics design space. We encourage future researchers to replicate and extend our work to examine how older adults use other crisis apps, and their perspectives on using such apps more broadly. While our analysis did not incorporate all concepts from Maslow's heirarchy of needs~\cite{mcleod2007maslow}, we encourage future theoretical, design, and empirical work investigating the varied levels of needs that arise in emergency scenarios, as well as how crisis apps should reenvisioned to better meet such needs for older adults.

\section{Conclusions}

Crisis apps can enable vitally important support during emergencies, such as information dissemination and the provision of assistance. However, these tools are usually focused on the general population, and do not consider the needs of subpopulations who are vulnerable, such as older adults. We hope that our work will help catalyze crisis informatics research that considers the specific needs of older adults. Given the increasing frequency of large  and destructive disaster events, there is a serious need for future work that examines how crisis apps can be designed to meet the needs of all, with a particular focus on empowering vulnerable populations.

\section{Acknowledgments}

This work was supported by Northeastern University's TIER 1: Seed Grant/Proof of Concept Program (cycle number FY19). We thank the reviewers for their constructive reviews and our colleagues for their feedback. We also thank our participants for sharing their stories and experiences with us, and the recruitment site for this research.
  
\bibliographystyle{SIGCHI-Reference-Format}
\bibliography{sample}

\end{document}